\documentclass[11pt]{article}
\usepackage{moriond,epsfig}

\def\Journal#1#2#3#4{{#1} {\bf #2}, #3 (#4)}

\def\NPB{{\em Nucl. Phys.} B}
\def\PLB{{\em Phys. Lett.}  B}

\def\ZPC{{\em Z. Phys.} C}
\def\EPJC{{\em Eur. Phys. J.} C}
\def\MPLA{{\em Mod. Phys. Lett.} A}
\def\APB{{\em Acta Phys. Polon.}B}

\def\be{\begin{equation}}
\def\ee{\end{equation}}
\def\bea{\begin{eqnarray}}
\def\eea{\end{eqnarray}}

\def\lsim{\raise0.3ex\hbox{$<$\kern-0.75em\raise-1.1ex\hbox{$\sim$}}}
\def\gsim{\raise0.3ex\hbox{$>$\kern-0.75em\raise-1.1ex\hbox{$\sim$}}}

\begin{document}
\vspace*{4cm}
\title{DIS AT LOW X, SATURATION SCALE, GLUON STRUCTURE FUNCTION AND
VECTOR-MESON PRODUCTION \footnote{Presented at 41st Rencontres de Moriond,
March 2006}}

\author{ D. SCHILDKNECHT }

\address{Fakult{\"a}t f{\"u}r Physik, Universit{\"a}t Bielefeld, D-33501 
Bielefeld, Germany\break
and\break
Max-Planck Institut f{\"u}r Physik (Werner-Heisenberg-Institut), 
F{\"o}hringer Ring 6, D-80805 M{\"u}nchen, Germany}

\maketitle\abstracts{Deep inelastic scattering at low x can be described
by essentially only two fitted parameters. The interpretation of
$J/\psi$ photoproduction in terms of the gluon structure function is 
elaborated upon.
}

I will concentrate on the intimate connection between the x-dependence
and the $Q^2$ dependence of the structure function $F_2 (x, Q^2)$, and
subsequently I will turn to vector-meson production, to $J/\psi$ production
in particular.

In deep inelastic scattering (DIS) at low $x \simeq Q^2/W^2 \ll 0.1$, the
photon fluctuates into a $q \bar q$ color-dipole state that in the
virtual forward-Compton-scattering amplitude interacts \cite{Nikolaev} via
the generic structure of two-gluon exchange with the proton. The QCD
gauge-theory structure implies diagonal and off-diagonal 
transitions \cite{Cvetic,Schi} in the masses of the color-dipole vector
states, and accordingly it implies a dependence on the transverse 
three-momentum of the gluon, $\vec l_{\bot}$, that couples to the color
dipole. The effective value of $\vec l_\bot$ introduces a novel scale,
the saturation scale, relevant in low-x DIS. In our approach, the 
saturation scale, $\Lambda^2_{sat} (W^2)$, depends on the energy, $W$, 
and \cite{Kuroda}
\be
\Lambda^2_{sat} (W^2) = \frac{1}{6} \langle \vec l^{~2}_\bot \rangle
\cong \frac{1}{6} {\rm const} \left( \frac{W^2}{1 GeV^2} \right)^{C_2}.
\label{eq: 1}
\ee
A fit to the total photoabsorption cross section by the power law (1) in
the HERA energy range gave \cite{Schi}
\be
2 GeV^2 \lsim \Lambda^2_{sat} (W^2) \lsim 7 GeV^2,
\label{eq: 2}
\ee
where
\bea
const & = & 0.340 \pm 0.063 GeV^2,\nonumber \\
C_2 & \equiv & C_2^{exp} = 0.27 \pm 0.01.
\label{eq: 3}
\eea
In addition to $\Lambda^2_{sat} (W^2)$, the total (virtual)
photoabsorption cross section depends on the cross section 
$\sigma^{(\infty)}$ of hadronic size,
\bea
&& \sigma^{(\infty)} = 48 GeV^{-2} = 18.7 mb,\\
&& ({\rm for}~R_{e^+e^-} = 10/3,~{\rm four~flavours}),\nonumber
\label{eq: 4}
\eea
and is approximately given by
\be
\sigma_{\gamma^*p} (W^2, Q^2) \cong \frac{\alpha}{3 \pi} R_{e^+e^-}
\sigma ^{(\infty)} \cdot \left\{ \begin{array}{ll}
\ln \eta^{-1}, & (\eta \ll 1),\\
\frac{1}{2} \eta^{-1}, & (\eta \gsim 1),
\end{array} \right. 
\label{eq: 5}
\ee
with the scaling variable \cite{Diff,Schi}
\be
\eta (W^2, Q^2) = \frac{Q^2 + m^2_0}{\Lambda^2_{sat} (W^2)}
\label{eq: 6}
\ee
and $m^2_0 \simeq 0.15 GeV^2$. Apart from this threshold mass,
the cross section (5), or equivalently $F_2(W^2, Q^2)$, contains
three adjusted parameters, the two parameters (3) determining the
saturation scale and the cross section (4).

Application of DGLAP evolution in the region of $Q^2 \gg \Lambda^2_{sat}
(W^2)$, where appropriate, actually reduces the number of three to
only two adjusted parameters, since evolution allows one to determine
the exponent $C_2$ in (1). This will be pointed out next.

The representation (5) of the experimental data contains the assumption 
that the scattering amplitude for longitudinal, $(q \bar q)^{J=1}_L$,
(vector) states and for transverse ones, $(q \bar q)^{J=1}_T$, be
proportional to each other. In terms of the sea-quark, $x \Sigma (x, Q^2)$,
and the gluon distribution, $xg (x,Q^2)$, and the proportionality constant
$r$, this proportionality reads \cite{Ku-Schi}
\bea
x \Sigma (x,Q^2) & = & \frac{12}{R_{e^+e^-}} F_2 (x, Q^2) = \frac{8}{3 \pi}
\left( r + \frac{1}{2}\right) \alpha_s (Q^2) xg (x, Q^2) \vert_{x=Q^2/W^2}\\
& = & \frac{1}{3 \pi^3} \left( r + \frac{1}{2} \right) \sigma^{(\infty)}
\Lambda^2_{sat} (W^2).\nonumber
\label{eq: 7}
\eea
The constant $r$ also determines the ratio of the longitudinal to the
transverse photoabsorption cross section,
\be
\frac{\sigma_{\gamma^*_L p} (W^2,Q^2)}{\sigma_{\gamma^*_T p} (W^2,Q^2)}
= \frac{1}{2r}.
\label{eq: 8}
\ee
The (successful) representation \cite{Schi} of the experimental data 
was based on
$r = 1$.
With (5) and (7), the evolution equation (at low x) \cite{Prytz}
\be
\frac{\partial F_2 (\frac{x}{2}, Q^2)}{\partial \ln Q^2} =
\frac{R_{e^+e^-}}{9 \pi} \alpha_s (Q^2) x g (x, Q^2)
\label{eq: 9}
\ee
turns into an equation for $\Lambda^2_{sat} (W^2)$. Inserting
the power law (1), one finds a constraint on $C_2$ that is given by 
\cite{Ku-Schi}
\be
(2r+1) 2^{C_2} C_2 = 1.
\label{eq: 10}
\ee

In Table 1, we show the relation between $r$ and $C_2$ resulting from
(10). The constant $r$, according to (7), determines the relative magnitude
of gluon
to sea distribution. The dependence of the structure function $F_2 (W^2)
= F_2 (Q^2/x)$ for $Q^2 \gg \Lambda^2_{sat} (W^2)$ follows from (5).
\newpage

\begin{table}[htb]
\caption{Results for $C_2^{theor.}$ for different values of $r$ according
to (10).\label{tab:tab1}}
\vspace{0.4cm}
\begin{center}
\begin{tabular}{|c|c|c|c|c|}
\hline
& & & & \\
r & $C_2^{theor.}$ & $\alpha_s \cdot {\rm glue}$ & 
$\sigma_{\gamma^*_L}/\sigma_{\gamma^*_T}$ & $F_2 \left( \frac{Q^2}{x} \right)$
\\ \hline
$\to \infty$ & 0 & $\ll {\rm sea}$ & 0 & $(Q^2/x)^0 = const.$ \\
1 & 0.276 & $\approx {\rm sea}$ & $ \sim \frac{1}{2}$ & $(Q^2/x)^{0.276}$ \\
0 & 0.65  & $> {\rm sea}$ & $\infty$ & $(Q^2/x)^{0.65}$\\ \hline
\end{tabular}
\end{center}
\end{table}

We summarize:
\begin{itemize}
\item[i)] The theoretical value of $C_2$ in Table 1 from (9) and (10)
for $r = 1$ coincides with the experimental one (3) obtained for $r = 1$,
\begin{displaymath}
C^{theor.}_2 \simeq C_2^{exp.},
\end{displaymath}
and thus the underlying ansatz for the dipole cross section is consistent
with the evolution equations from QCD. A (strong) violation of (10) would
have ruled out this ansatz, and in particular the underlying assumption
of $W$ being the relevant variable to describe diffractive processes at
low $x$.
\item[ii)] Essentially two parameters, the normalization of the saturation
scale $\Lambda^2_{sat} (W^2)$ in (3) and the cross section of hadronic
magnitude (4) are sufficient to determine the low-$x$  proton structure
function including the photoproduction limit.
\item[iii)] The $Q^2$ and the $x$ dependence of $F_2 (x, Q^2)$ are
strongly correlated with each other and correlated with the relative 
magnitude of the gluon and sea contributions, compare Table 1.
\item[iv)] A sufficiently large gluon contribution implies a strong rise
of $F_2 (x, Q^2)$ with increasing $Q^2$ for constant $x$, and an equally
strong rise with decreasing $x$ at fixed $Q^2$ (compare lines 2 and 3 in
Table 1). This qualitative feature is experimentally realized, and 
theoretically it is a natural consequence of $W$ as the relevant 
variable that describes the scattering cross section of a color dipole
on the proton (rather than $x$).
\item[v)] Since the relative magnitude of the gluon and the sea is
correlated with $\sigma_{\gamma^*_Lp}/\sigma_{\gamma^*_Tp}$, direct
measurements of this ratio are urgently needed. This allows one to
investigate the limits of validity of the underlying assumed proportionality
of sea and gluon distributions.
\end{itemize}
Turning to $J/\psi$ production, in figs. 1 and 2, I show our result
\cite{Schi-Ku,Kuroda} of an absolute prediction based on the description of
the inclusive DIS data I told you about. For details, I have to refer
to the original publications.

I wish to mention one important point, however, related to the interpretation
of $J/\psi$ photoproduction $(Q^2 = 0)$ in terms of the gluon structure
function. From (7), valid for sufficiently large $Q^2 \gg \Lambda^2_{sat}
(W^2)$, we have
\be
\alpha_s (Q^2) x g (x, Q^2) \vert_{x=Q^2/W^2} = \frac{1}{8 \pi^2}
\sigma^{(\infty)} \Lambda^2_{sat} (W^2 = Q^2/x).
\label{eq: 11}
\ee
According to (11), a determination of the energy dependence of
$\Lambda^2_{sat} (W^2)$ at any $Q^2$, e.g. at $Q^2 = 0$ in $J/\psi$
photoproduction, yields the dependence of the gluon structure function
on the left-hand side as a function of $x$ at $Q^2 \gg \Lambda^2_{sat}
(W^2)$, where relation (11) becomes valid. Clearly, the measurement
of $J/\psi$ photoproduction does {\it not} provide a measurement of
the structure function for $Q^2 \lsim m^2_c, \Lambda^2_{sat} (W^2)$,
where (11) breaks down.

\begin{figure}[ht]
\setlength{\unitlength}{1cm}
\begin{center}
\begin{minipage}[t]{7cm}
\begin{picture}(6.5,6.5)\psfig{file=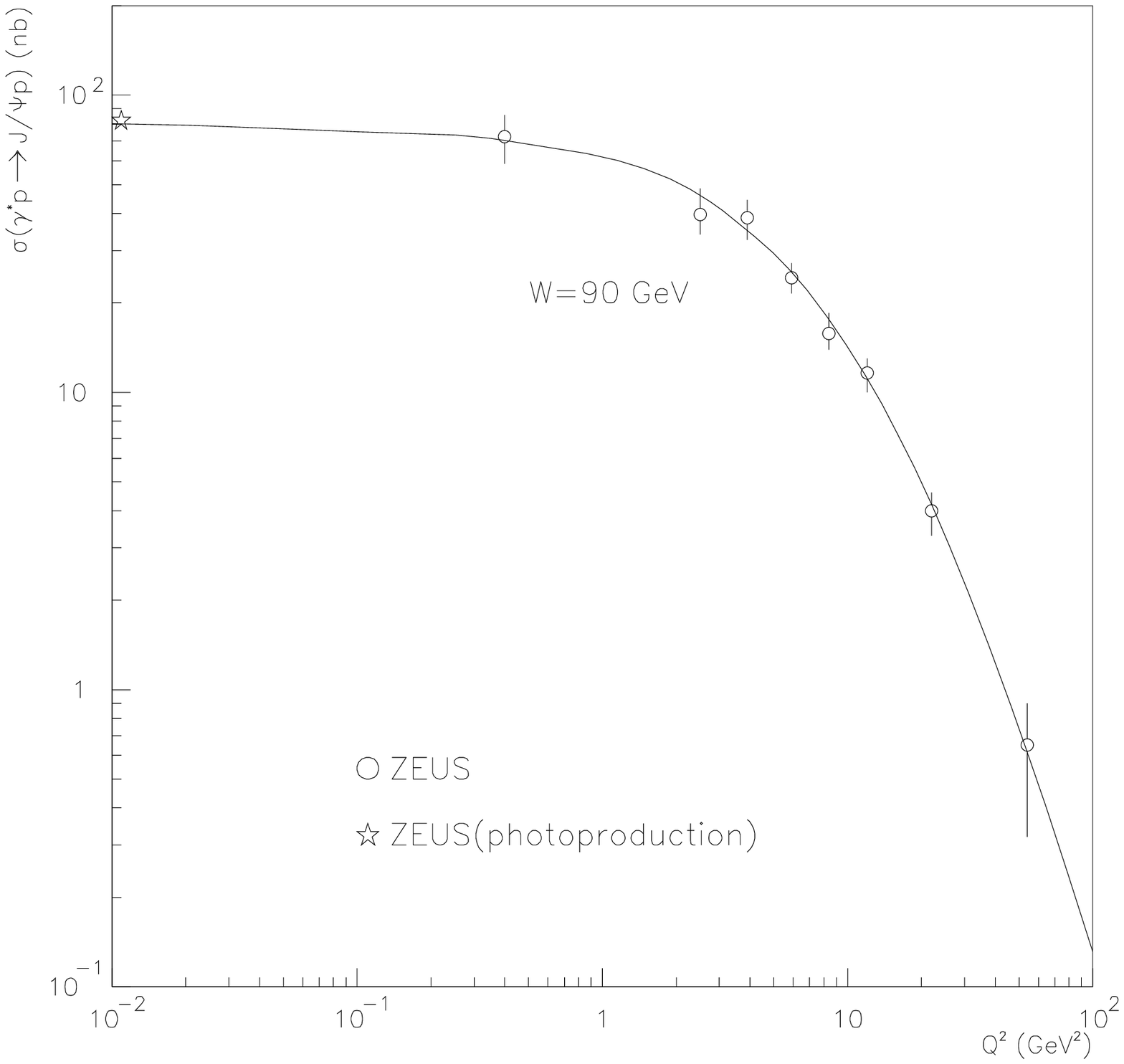,width=7cm,height=7cm}
\end{picture}\par
\caption{The $Q^2$ dependence of the cross section for $J/\psi$ production.}
\end{minipage}\hspace*{1cm}
\begin{minipage}[t]{7cm}
\begin{picture}(6.5,6.5)\psfig{file=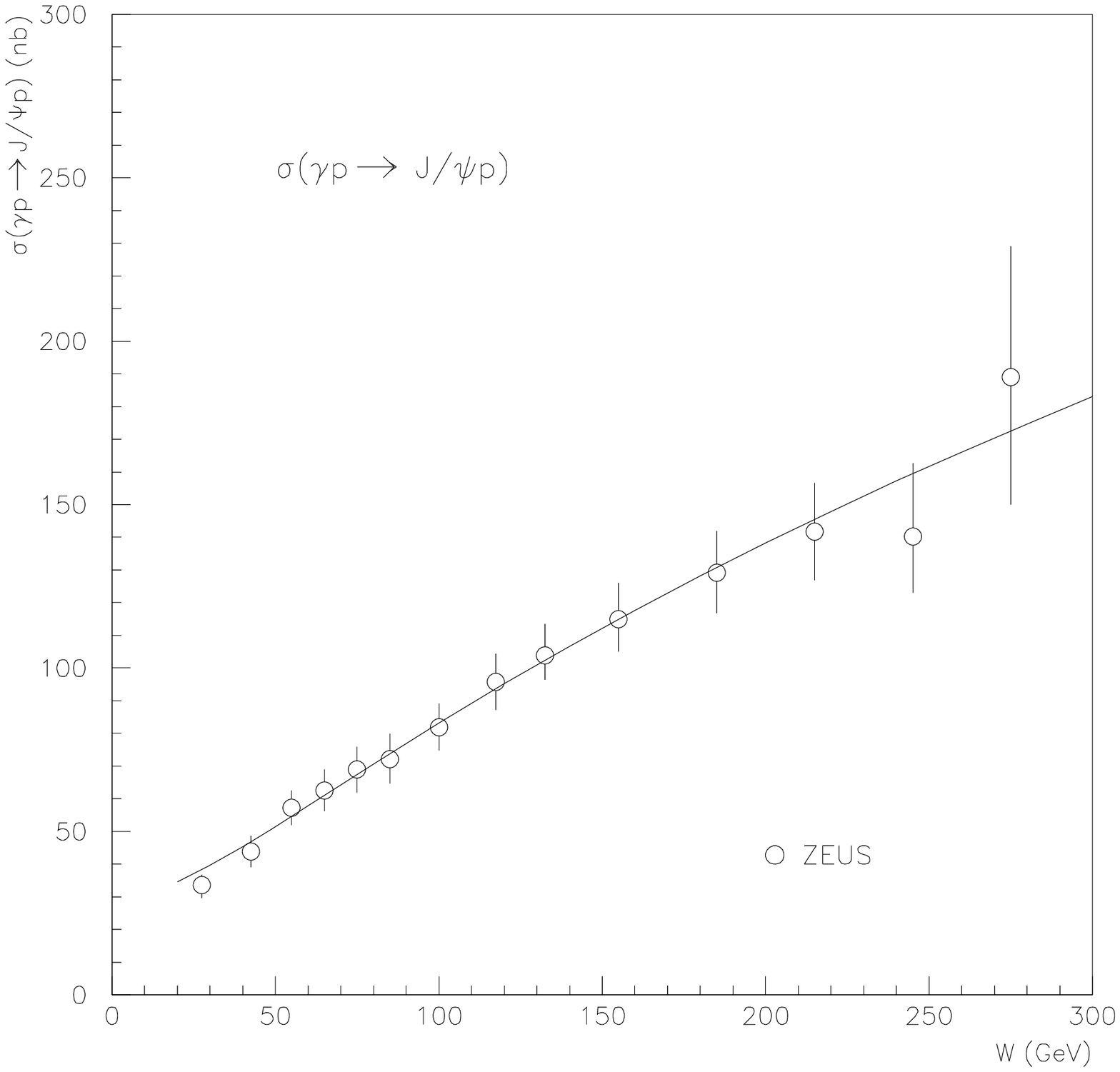,width=7cm,height=7cm}
\end{picture}\par
\caption{The $W$-dependence of $J/\psi$ photo-production}
\end{minipage}
\end{center}
\end{figure}

More generally, independent of {\it our} representation of the data on DIS, 
{\it any}
unique prediction of $J/\psi$ photoproduction necessarily requires 
the left-hand side of (11) to only depend on $W^2$. Otherwise no unique
prediction of $J/\psi$ photoproduction will emerge. This should be kept
in mind, when predicting the energy dependence of vector meson 
photoabsorption, i.e. for {\it any} specific fit of the gluon structure 
function
the left-hand side in (11) should be examined on whether it only
depends on $W^2$ in good approximation at large $Q^2$.

\section*{Acknowledgements}
Thanks to Masaaki Kuroda for a fruitful collaboration.\hfill\break
This work was supported by Deutsche Forschungsgemeinschaft under
grant SCHI 189/6-2.

\end{document}